\documentclass[twocolumn]{aastex63}

\newcommand{\civ}{\ion{C}{4}}

\received{}
\revised{}
\accepted{}
\submitjournal{ApJL}

\shorttitle{Quasar Orientation}
\shortauthors{Richards et al.}

\graphicspath{{./}}

\begin{document}

\title{A Novel Test of Quasar Orientation}

%% you desire. In v6.3 you can include a footnote in the title.

%\correspondingauthor{August Muench}
%\email{greg.schwarz@aas.org, gus.muench@aas.org}

\author[0000-0002-1061-1804]{Gordon T. Richards}
\affiliation{Department of Physics, Drexel University, 32 S.\ 32nd Street, Philadelphia, PA 19104, USA}

%%% N.B.  Any of the authors below could end up 2nd author
%%% depending on what actually makes it into the paper
\author[0000-0002-7092-0326]{Richard M. Plotkin}
\affiliation{Department of Physics, University of Nevada, Reno, NV 89557, USA}

\author[0000-0002-6528-1937]{Paul C. Hewett}
\affiliation{Institute of Astronomy, University of Cambridge, Madingley Road, Cambridge, CB3 0HA, UK}

\author[0000-0002-2091-1966]{Amy L. Rankine}
\affiliation{Institute of Astronomy, University of Cambridge, Madingley Road, Cambridge, CB3 0HA, UK}

\author[0000-0001-8125-1669]{Angelica B. Rivera}
\affiliation{Department of Physics, Drexel University, 32 S.\ 32nd Street, Philadelphia, PA 19104, USA}

\author[0000-0003-1659-7035]{Yue Shen}
\affiliation{Department of Astronomy, University of Illinois at Urbana-Champaign, Urbana, IL 61801, USA}
\affiliation{National Center for Supercomputing Applications, University of Illinois at Urbana-Champaign, Urbana, IL 61801, USA}

\author[0000-0003-4327-1460]{Ohad Shemmer}
\affiliation{Department of Physics, University of North Texas, Denton, TX 76203, USA}

\begin{abstract}

%Currently 131 words, can be as long as 250.
The orientation of the disk of material accreting onto supermassive black holes that power quasars is one of most important quantities that are needed to understand quasars---both individually and in the ensemble average.
We present a hypothesis for determining comparatively edge-on orientation in a subset of quasars (both radio loud and radio quiet).  If confirmed, this orientation indicator could be applicable to individual quasars without reference to radio or X-ray data and could identify some 10--20\% of quasars as being more edge-on than average, based only on moderate resolution and signal-to-noise spectroscopy covering the \civ\ $\lambda$ 1549 \AA\ emission feature. We present a test of said hypothesis using X-ray observations and identify additional data that are needed to confirm this hypothesis and calibrate the metric.

\end{abstract}

%\keywords{editorials, notices --- miscellaneous --- catalogs --- surveys}

%\keywords{quasars: general -- quasars: emission lines}
%GTR: No longer used.  Instead use UAT upon submission.
%Radio quiet quasars(1354)
%Radio loud quasars(1349)
%Optical observation(1169)
%Emission line galaxies(459)
%X-ray quasars (1821)
%Metal line absorbers (1032_

\section{Introduction} \label{sec:intro}

The physics of quasars and active galactic nuclei (AGNs) is primarily governed by three properties of the
system: the mass of the black hole, the spin of the black hole, and the
accretion rate---with our line-of-sight orientation to the accretion
disk also strongly affecting how we see and interpret these systems.  Astronomers have gone to great lengths to measure black hole
masses for the $\approx$100 quasars with robust ``reverberation mapping"
analysis \citep{Peterson1993} and have constructed empirical ``scaling relations" \citep[e.g.,][]{Vestergaard+2006} that allow astronomers to estimate black hole masses for a much larger sample
of AGN/quasars---albeit possibly biased toward objects that lack
evidence for strong accretion disk winds \citep[e.g.,][]{Richards+2011,Shen2013}.

The situation for black hole spins is much worse.  Using measurements of
subtle signatures of gravitational redshifts from atomic features
present in X-ray data, it has been possible to estimate black hole spin for
only $\approx$2 dozen AGNs \citep{Reynolds19}.  Similarly, the ``beam power" of the outflow in strong radio sources has been used in attempt to estimate the
spin for only 55 AGNs \citep{Daly11}.  So-called ``thermal continuum"
fitting procedures \citep[e.g.,][]{Capellupo+15}
%Capellupo+16
have shown some promise, but with a high degree of uncertainty and also limited
to handfuls of AGNs.

The third parameter, the accretion rate, is historically estimated
from the quasar luminosity as $L=\eta\dot{M}c^2$, with the
uncertainties inherent to understanding how a monochromatic luminosity
translates to a bolometric luminosity %\citep[e.g.,][]{Krawczyk+2013} 
and the aforementioned spin dependence that affects $\eta$.

Thus, it may be that orientation (while not fundamental to quasar physics itself, but nevertheless fundamental to our ability to understand said physics) is the best measured of these key AGN parameters.
Specifically, if we adopt a model where all quasars have essentially
the same axisymmetric geometry (with the accretion disk
obscured by a dusty toroidal region when observed edge-on, e.g., \citealt{Elitzur12}), then we have thousands of examples of so-called ``type-2" quasars that are almost certainly observed edge-on (e.g., \citealt{Zakamska+03,Reyes+08}).  However, for type-2 quasars, {\em all} of the crucial diagnostics that can be derived from the accretion disk (such as the black hole mass) are gone, rendering the orientation information rather emasculated.  Alternatively, it has long been argued that radio
spectral index provides a rough orientation estimate for quasars
\citep{OB82,rlb+01,vwr+15}, but only $\approx$5\% of
quasars from the Sloan Digital Sky Survey (SDSS; \citealt{York+2000}) are even radio-detected by the moderately deep and large-area Faint Images of the Radio Sky at Twenty centimeters (FIRST; \citealt{FIRST}) survey \citep{Ivezic+02,KR2015}.  Moreover, most radio emission from quasars may have little to do with jets \citep[e.g.,][]{Panessa+19}. %Kimball

In this Letter we hypothesize another orientation measure for quasars that has the potential to be applicable for both radio-loud and radio-quiet quasars.
%possibly for more than just 5\% of the quasar population.  
In Section~\ref{sec:hypothesis} we describe the proposed metric and an experiment that can be applied to test the metric.  In Section~\ref{sec:data} we present the data for this test.  We carry out a test of a prediction in Section~\ref{sec:test} and identify shortcomings that suggest the need for additional data.  We finish with discussion and conclusions in Section~\ref{sec:conclusions}.

\section{The Hypothesis}
\label{sec:hypothesis}
 
In terms of accurately determining one of the key parameters needed to understand the physics of quasars, we argue herein that a promising avenue is the use of absorption and emission features in quasars to identify broad-line quasars with comparatively edge-on line-of-sight orientations\footnote{Hereafter we shall use ``edge-on" to mean as edge-on as possible without obscuring the accretion disk continuum and the broad-line region.}.
Our proposed orientation indicator is the presence of \civ\ absorption at the systemic redshift of the quasar (which is often {\em redward} of the peak of the \civ\ emission line).  The background needed to understand the orientation hypothesis for such systems 
%follows.  Through careful analysis and machine-learning techniques it has become possible to determine 
starts with determining the redshifts of quasars very accurately \citep[e.g.,][]{Hewett+2010}, removing much of what can be thousands of kilometers per second of uncertainty due to accretion disk winds \citep{Richards+02,Dix+20}.
%\citep{VandenBerk+01,Richards+02}.  
As a result of that process, it has become clear that what was once thought to be a distribution of \civ\ absorption line systems representing ``cluster" gas with velocities within $\pm 1000\,{\rm km\,s^{-1}}$ of the systemic redshift \citep{Foltz+1986,Anderson+87} are more likely a combination of outflowing material and ``virialized" material that is at much lower velocity \citep{Bowler+14}.
%(if not identically zero); 
%see \cite{Bowler+14} and \citet{Stone+2019}.  
%Indeed, the appearance of a \civ\ absorption system {\em redward} of the peak of the \civ\ line may be one of the best indicators of true systemic redshifts.  
\citet{Stone+2019} argued that systems with ``zero-velocity associated absorption-line systems" (hereafter AAL0s), may be an orientation indicator for radio-quiet quasars given that (1) in radio-loud quasars these systems are observed preferentially in steep-spectrum radio sources that are presumed to have an edge-on orientation, and (2) that they are just as common in radio-quiet quasars as in radio-loud quasars.

Given the prevalence of \civ\ absorption due to gas in the halos of galaxies \citep[e.g.,][]{Chen+01,Prochaska+14,Perrotta+16}---including our own Milky Way
%HP07 (Hennawi)
\citep[e.g.,][]{Richter+17}---it might be expected that one should see \civ\ in absorption at $\approx$0 velocity in nearly every galaxy unless something has happened to that gas.
%[Steidel1990?, Churchill1999?, another MW paper], 
One reason for said gas not to be there (and causing absorption at the appropriate ionization) is if quasar activity has driven the gas from our line of sight.  Arguably, the most likely place for such gas to remain along our line of sight is in a region that is relatively shielded from the accretion disk (and the wind blown from it; e.g., \citealt{Giustini+2019}), where gas in the halo of the host galaxy might persist.  
Alternatively, \citet{Ganguly+01} suggested that AALs may represent gas that is ``hugging" an equatorial accretion disk wind (likely to
be seen more edge-on than not).
Thus, whether the gas is associated with the central engine or the host galaxy, AAL0s might be expected to be more common in relatively edge-on orientations.  Such an explanation for AAL0s would be consistent with the findings of \citet{Stone+2019} and could provide an orientation indicator for both radio-loud and radio-quiet quasars that exhibit AAL0s.

During the course of an analysis of repeat spectroscopy of quasars
from the SDSS Reverberation Mapping (SDSS-RM; \citealt{SDSSRM}) campaign, \citet{Rivera+20}
%Shen+2019
discovered many examples of quasars with AAL0s (10--20\%), which potentially   
%The fraction of such quasars does not appear to be as large as the fraction of quasars that exhibit BAL troughs \citep[e.g.,][]{Allen+11}; however, it could 
represent a larger fraction of quasars than current estimates of orientation allow.   For reference, a symmetric conical distribution of absorbing gas with a 10--20\% covering fraction would correspond to a opening angle of 6--12 degrees with respect to the plane of the accretion disk (in the absence of toroidal obscuration).
If AAL0 systems are indicative of edge-on orientation, the SDSS-RM sample would be a unique sample for an investigation of orientation given the abundance of data on these systems.  Specifically, while not all of the SDSS-RM quasars have accurately estimated black hole masses, we can be certain that the time-dependent changes mapped by the SDSS-RM campaign are not caused by changes in the black hole mass and are unlikely to be caused by orientation.  

Our hypothesis is that all quasars with AAL0 systems have edge-on orientations, and we suggest a novel test of this hypothesis using X-ray data.  This test applies only to a fraction of the AAL0 quasars; however, positive confirmation of an edge-on orientation for this subsample would add confidence to the edge-on hypothesis for the parent sample.  

Specifically, in \citet{Wu+11}, \citet{Luo+2015} and \citet{Ni+2018}, it was argued that a ``slim"
accretion disk model \citep{Abramowicz1988} might be able to explain the
observed diversity of X-ray properties of weak-lined quasars (WLQs).  WLQs are quasars where Ly$\alpha$+\ion{N}{5} EW $\lesssim10$--15~\AA\  or \civ\ EW $\lesssim 10$~\AA\ \citep{Shemmer+09,Diamond+09}.
%[Other references to consider: Fan99,Anderson01,Colling05,Fan06,Plotkin+10,SL15,Plotkin+15].  
Figure~18 from \citet{Luo+2015} illustrates the empirical problem and its proposed solution.  The problem is that WLQs are observed to have X-ray measurements that are sometimes normal (relative to their UV luminosity, given the well-established $L_{\rm UV}-\alpha_{\rm ox}$ relationship), but sometimes are X-ray weak.  The \citet{Luo+2015} model would argue that {\em all} of the WLQs are hosted by quasars with slim disks that shield the broad-emission-line-region (BELR) gas from being over-ionized by X-ray radiation from the hot corona (see also \citealt{Giustini+2019} for a more detailed illustration of how wind strength depends on accretion rate and black hole mass).  While the BELR gas is always shielded from the X-ray corona in such slim-disk systems, only more edge-on WLQs would be X-ray weak from our line of sight.  For face-on orientations, an Earth-based observer will see the X-ray corona even if the BELR does
not.  Thus an X-ray indicator of orientation is possible: the geometry of the accretion disk creates a situation conducive to forming WLQs, but the orientation of that disk with respect to our line of sight dictates whether we see such objects as X-ray weak or X-ray normal.  It is therefore an important step forward in determining quasar orientation to observe samples of WLQs in the X-ray.  Indeed a number of investigations have done just that \citep[e.g.,][]{Shemmer+09,Shemmer+10,Wu+11,Luo+2015,Ni+2018,Marlar+18}.  

If AAL0s are indicative of edge-on orientation and if X-ray weakness in quasars is also evidence for edge-on orientation in objects with slim accretion disks, then we might expect both of the following to be true: (1) quasars with AAL0s are more likely to be X-ray weak if they have slim disks, and (2) quasars with slim disks are more likely to be X-ray weak if they have AAL0s.  While these two situations sound similar, they are indeed distinct as the first asks about the X-ray properties of all quasars purported to be edge-on (regardless of whether they host a slim disk) and the second asks about quasars with slim disks (regardless of whether they are purported to be edge-on or not.)

\section{Data}
\label{sec:data}

%Need SDSSRM b/c can see lower EQW AAL0s.

The SDSS-RM program
%Reverberation Mapping program (SDSS-RM; \citealt{SDSSRM}) 
took dozens of epochs of repeat spectroscopy
%Shen+2019
of 849 broad-line quasars in a 7 square degree field of view.  This experiment was designed to determine the time delay between continuum and emission-line variations in order to estimate black hole masses using the reverberation mapping technique.
%\citep{Peterson1993}.  
\citet{Rivera+20} analyzed the spectra of 133 of these 849 quasars that have 30 or more epochs, with a mean S/N$>6$ per pixel over the wavelength interval $3650 < \lambda < 9300$\,\AA. 
%included in the SDSS DR14 release.  
These data enabled an investigation of how much limited spectrum S/N and intrinsic spectral variability can affect the measurement of \civ\ equivalent width (EW) 
and \civ\ blueshift  (see also \citealt{Sun+2018}), which are thought to be indicators of accretion disk winds \citep{Richards+2011}.  As there is a degeneracy between \civ\ EW and blueshift 
%in terms of determining extrema in terms of \civ\ emission-line properties 
(e.g., quasars with intermediate blueshift or EW can have a large range of the other property; \citealt{Richards+2011}),  \citet{Rivera+20} define a hybrid metric, the \civ\ ``distance", as a more robust wind indicator.  This distance is relative to the best-fit curve tracing the locus of points in the \civ\ EW--blueshift plane, where quasars with large EW and small blueshift have small distance, while quasars with small EW and large blueshift have large distance.  %This metric should be more robust in terms of identifying quasars with slim-disk-like geometries than EW or blueshift alone.
%; see also \cite{Sun+2018}.  

The \citet{Rivera+20} analysis was based on independent component analysis (ICA) reconstructions of the SDSS-RM spectra using the spectral components determined from \citet{Rankine+2020}.
As a result of those spectral reconstructions, these 133 quasars have high-S/N composite spectra and accurately determined continua, including in the region of the broad emission line.  These data are conducive to measuring both broad and narrow absorption features in the quasar spectra.
%, including weaker absorption features than can be probed by single-epoch SDSS spectroscopy.  
Because the ICA reconstruction process both requires and enables accurate determination of the systemic redshift (uncertainty $\approx$230 km s$^{-1}$ as compared to the $\approx1000$ km s$^{-1}$ from SDSS-I/II single-epoch spectra), it is possible to recognize absorption systems that are {\em redward} of the emission-line peaks despite being at/near the systemic redshift \citep{Bowler+14}.  We illustrate the \civ\ region of 15 spectra with ICA-based \civ\ blueshift greater than 1500\,km s$^{-1}$ in Figure 1. 

We find that 20--27 of the 133 quasars analyzed by \citet{Rivera+20} have AAL0s,
depending on the criteria chosen for equivalent width of the absorption feature and resolution/ionization (which may distinguish narrow, but deep broad absorption troughs [``mini-BALs"] at relatively small distances from the black hole from AALs, potentially at large distances).
We require the \civ\ doublet to be visually resolved and within $\approx \pm 500$ km s$^{-1}$ of the systemic redshift.  More work is needed to formally distinguish AAL0 systems from mini-BALs, so we adopt an AAL0 fraction of 10--20\% and include only the 20 most likely sources in our analysis herein.

%/Users/gtr/Dropbox/SDSSRMspec/SDSSRMdata/AAL0plot/AAL0plot.ipynb

\begin{figure}[tbh]
\epsscale{1.0}
%\plotone{SDSSRMAAL0s3panel.pdf}
\plotone{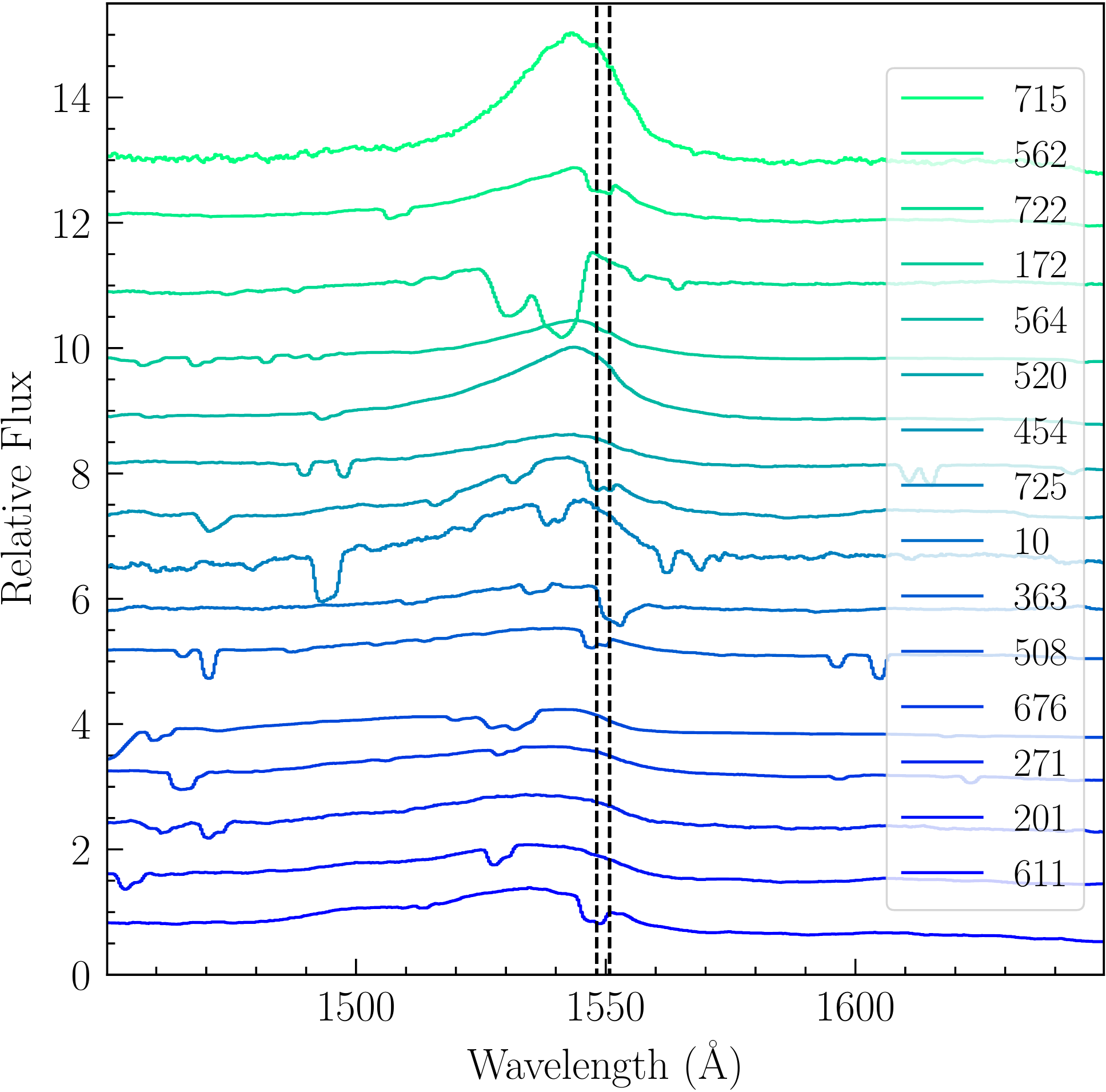}
\caption{\civ\ emission-line regions of the 15 highest blueshift sources in our sample.  Shown are the median spectra 
(from all of the SDSS-RM epochs).
%for each of these objects.  
Five have AAL0s (RMIDs 10, 363, 454, 562, and 611) and are candidates for edge-on orientation.  The vertical lines indicate the \civ\ doublet at $\lambda\lambda$ 1548.202, 1550.744.  RMID 722 is a BAL that should be correctly located in terms of \civ\ blueshift due to the ICA reconstruction process, but is expected to be absorbed in the X-ray.  None of these sources are formally WLQs, but the combination of large blueshift and low EQW (for most of the sources shown) are potentially indicative of a slim-disk-like geometry.
\label{fig:fig1}
}
\end{figure}

In addition to a sample where AAL0s can be recognized, our proposed test of the orientation hypothesis requires sensitive X-ray data.  The SDSS-RM field has 6.13 square degrees of X-ray coverage from {\em XMM-Newton} to an effective depth of $\approx 15\,$ks \citep{Liu+20}, with detection of 584 of the 849 quasars in the full SDSS-RM sample and 96 of the 133 quasars from \citet{Rivera+20}.  In addition, \citet{Liu+20} performed forced photometry at the location of the full SDSS-RM sample, which provides 2$\sigma$ upper limits for another 19 sources from \citet{Rivera+20}.

There is also archival multiwavelength coverage of the field.  
We include {\em Chandra} detections of RMIDs 452, 573, and 611 from the \textit{Chandra} Source Catalog Release (CSC) 2.0 \citep{CSC}, in addition to four upper limits from the CSC and four matches in the AEGIS field (\citealt{Nandra+15}, including another three that also have XMM-RM observations).  {\em Chandra} limits are taken as the sensitivity limits provided by the CSC for a `marginal’ detection at the location of the source.  

Thus we have X-ray detections or upper limits for $96+19+3+4+4=126$ of the sources analyzed in \citet{Rivera+20}, including 18 of the 20 AAL0s.

%#Let's take stock of what we have in terms of coverage of the 133 %ICA SDSS-RM sources
%print(len(XMMRMICAdata['RMID_1']),np.sort(XMMRMICAdata['RMID']))
%print(len(XMMRMICAforceddata['RMID_1']),np.sort(XMMRMICAforceddata['RMID']))
%print(len(AEGISICAdata['RMID_1']),np.sort(AEGISICAdata['RMID_1']))

%96 [ 11  31  41  45  51  59  75  81  87 113 116 117 119 124 130 153 159 161 172 178 180 201 202 205 207 227 237 238 253 262 271 275 283 298 299 311 317 321 327 330 342 343 348 353 359 363 372 381 387 401 405 408 416 424 435 451 454 456 461 474 485 493 507 508 517 520 540 543 554 555 561 563 564 578 591 609 613 616 623 630 635 676 688 692 693 710 715 718 737 770 774 784 796 811 818 831]
%19 [ 10  19  34  36 128 155 230 339 357 361 374 379 380 535 562 717 724 725 780]
%7 [130 142 195 220 227 231 299]

%The first 2 shouldn't have any overlaps and have 96 and 19 sources respectively.  AEGIS has 4 unique sources:
%142, 195, 220, 231, 
%and 3 overlaps:
%130, 227, 299

%Plotkin then adds detections of 
%452, 573, 611
%and limits on 
%671, 738, 743, and 765

%Lastly, we add a limit on
%520
%based on the flux of the nearest neighbor.  It is not clear why that object is not in the "forced" catalog.

%So, this is 96+19+4+3+4+1 = 127

%Using V9, AEGIS and forced photometry, I now find that 14 are missing
%[389, 409, 452, 524, 573, 611, 660, 671, 713, 722, 729, 738, 743, 765]
%unclear why 389, 409, 671 missing from XMM
%573, 660 on edge of XMM
%611 from Plotkin archival search
%Note that 520 had to be special cased due to lack of Flux_S in table.
%Of the missing, 524, 729 and 738 are strong AAL0s

\section{The Test}
\label{sec:test}

\begin{figure*}[tbh]
\epsscale{1.0}
%\plotone{ChandraCycle22_AAL0_strong_medianc4dst_v9_AEGIS.pdf}
\plotone{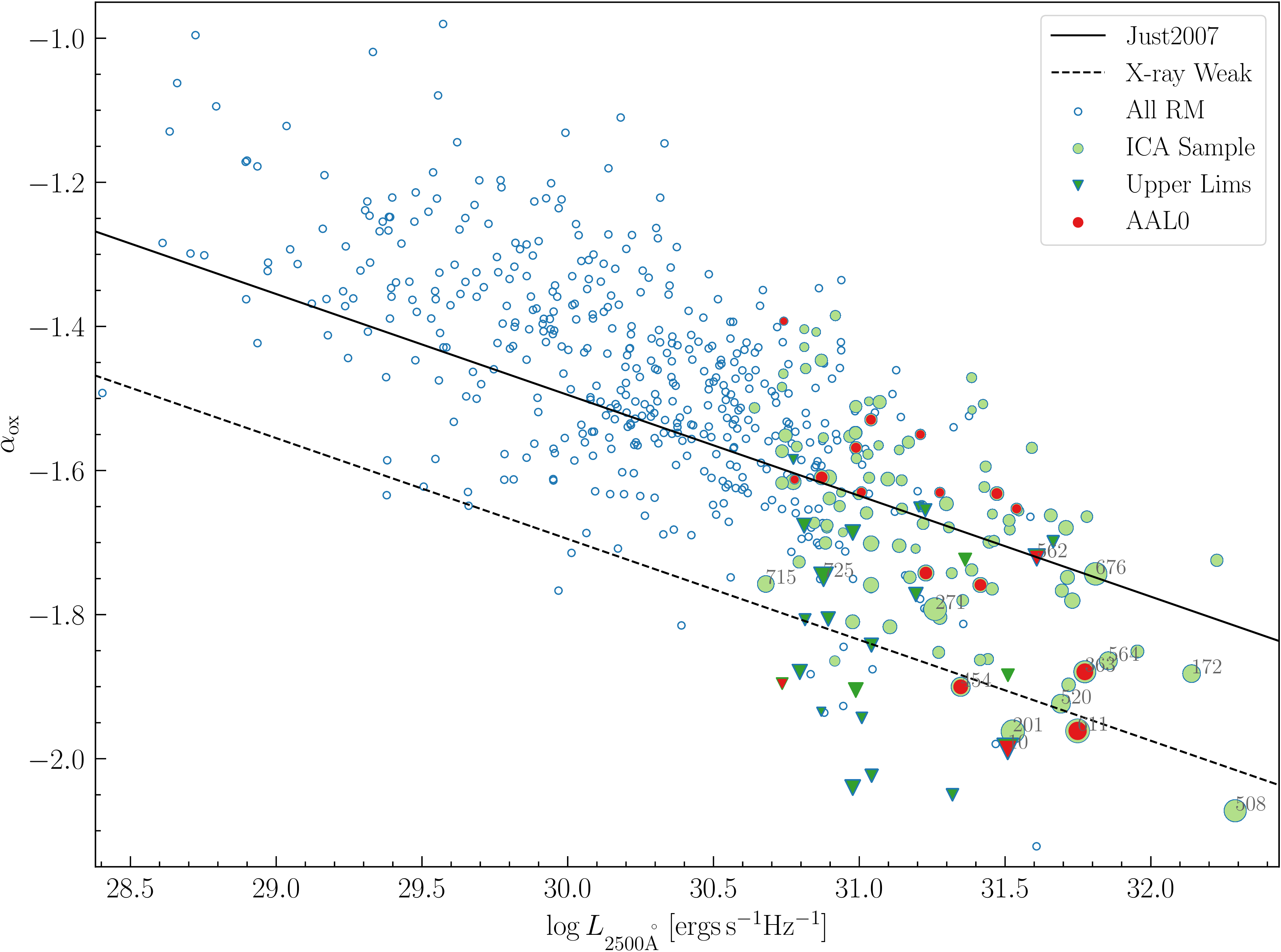}
\caption{Optical-X-ray flux ratio, $\alpha_{\rm ox}$ vs.\ UV luminosity, $\log L_{\rm 2500\AA} [{\rm ergs\,s^{-1} Hz^{-1}]}$ for the SDSS-RM sample.  Open blue points are all the SDSS-RM sources detected in the XMM-RM catalog and are plotted with uniform marker size.  The remaining marker types are all scaled by the ``distance" along the \civ\ parameter space: larger points have larger \civ\ distance (14 sources with the largest blueshifts also being labeled in grey by their SDSS-RM ID numbers).  Filled points are the sources investigated by \citet{Rivera+20}: light green indicates X-ray detections, and dark green inverted triangles are nondetections in the X-ray (2$\sigma$ upper limits).  Points include both data from XMM-RM and from archival observations.  %The pink diamond represents SDSS-RM ID 611, which has archival {\em Chandra} data.  
Sources with AAL0 systems (18 of the 20 identified by \citealt{Rivera+20} having X-ray data) are filled in red.  Our hypothesis predicts that larger points filled with red are more likely to be X-ray weak (as indicated by the dashed line relative to the best fit from \citealt{Just+07}).  
%[18 of 20 strong AAL0 sources have X-ray]  [14 of 15 high blueshift sources have X-ray] 
\label{fig:fig2}}
\end{figure*}

Using these data, Figure~\ref{fig:fig2} presents evidence that is consistent with the second prediction from \S~\ref{sec:hypothesis}, namely, that quasars with evidence of accretion disk winds are more likely to be X-ray weak if they host AAL0s.
%In Figure~\ref{fig:fig2} we see that this is indeed the case.  
We plot $\alpha_{\rm ox}$ vs.\ $L_{\rm UV}$  for the full SDSS-RM sample detected in the X-ray as open blue circles.   We highlight the subsample analyzed by \citet{Rivera+20} using filled light green circles for X-ray detections and dark green inverted triangles for (2$\sigma$) upper limits.  Sources with AAL0s are filled with red.  To indicate likelihood of having a strong accretion disk wind, the size of the points is scaled by the \civ\ distance, with the 14 largest blueshift objects indicated by their SDSS-RM IDs in grey next to the data point.  A dashed line indicates a luminosity-corrected X-ray weakness ($\Delta \alpha_{\rm ox}$ of $-0.2$; \citealt{Just+07}).

%/Users/gtr/Work/proposals/chandra/cycle22/cycle22targets.ipynb

Below $\log L_{2500{\rm \AA}} = 31.2$, we find only a single X-ray weak AAL0 source (of seven total) and that lone object corresponds to RMID 738 where the absorption is more suggestive of a mini-BAL than an AAL0.  Contrarily, at higher luminosity, we find that four of the five AAL0 objects with large \civ\ blueshift (RMIDs 10, 363, 454, and 611) are well below the best-fit $L_{\rm UV}$--$\alpha_{\rm ox}$ line and might be described as X-ray weak, thus confirming the second prediction of our hypothesis.  The fifth source (RMID 562) has $\Delta \alpha_{\rm ox} \approx 0$, but is an X-ray upper limit, so could be X-ray weaker than indicated.  
%These empirical findings would appear to provide confirmation of the viability of our hypothesis.

However, our current analysis is degenerate in a way that keeps us
from confirming the first prediction of the orientation hypothesis from $\S~\ref{sec:hypothesis}$.
Specifically, the data do not reveal a difference between the X-ray properties of luminous quasars with large blueshifts that do and do not host AAL0s.  It could be that the apparent confirmation of the second prediction is simply the result of blueshift being correlated with orientation or the known decrease in X-ray strength with increasing \civ\ blueshift \citep[e.g.,][]{Richards+2011}.  
This result may not be unexpected given that \citet{Ni+2018} find that there is a strong transition to X-ray weakness as the \civ\ emission-line EW goes
from 20 to 10\AA\
%and in \citet{Luo+2015} and \citet{Ni+2018} the indicator of a slim disk is a WLQ as they are expected to have high mass-weighted accretion rates ($L_{\rm bol}/L_{\rm Edd}$, the Eddington ratio).  Our 
and our sources have a minimum \civ\ EW of $\approx 10$\AA.  Thus our sample, lacking formal WLQs, 
%of the type that \citet{Luo+2015} argue may exhibit slim disks, 
may not be a good test of the orientation hypothesis.
% as we would not expect to see a large fraction of X-ray weak
%AAL0s.  

Nevertheless, we note the well-known rise of median \civ\ blueshift with UV luminosity \citep[e.g.,][]{Richards+2011} and the suggestion that winds may require $L_{\rm bol} > 3\times10^{45}\, {\rm ergs\,s^{-1}}$ \citep[e.g.,][]{Veilleux+2013, Zakamska+2014}, which corresponds to $\log L_{\rm 2500\AA} \approx 30$---that is, lower than the proposed slim-disk systems herein.  Moreover, there is no evidence for a sharp phase transition in the distribution of \civ\ emission-line blueshifts, so it seems likely that the development of a slim disk would be a smooth transition (e.g., in terms of scale height for X-ray absorption) rather than an abrupt one.   Thus, high-luminosity, high-blueshift SDSS-RM sources still are likely to host a strong accretion disk wind, which may indicate a similar, if less extreme, accretion disk geometry.  %Specifically, high UV luminosity (needed to drive the wind) and large \civ\ emission-line blueshifts (as an indicator of wind strength) provide a potential indication of a slim-disk-like geometry (as do low \civ\ EW or high Eddington radio)}---for example in a model where that geometry is needed to shield the emission line gas from overionization and allow effective radiation line driving.
Indeed, the WLQ sample of \citet{Luo+2015} is predominantly sources with \civ\ blueshifts in excess of 2000\,km s$^{-1}$ (where we adopt a positive, rather than a negative sign convention to represent an outflow).  
%As there is a degeneracy between \civ\ EW and blueshift in terms of determining extrema in terms of \civ\ emission-line properties (quasars with the intermediate blueshift and EW can have a large range of the other property, \citealt{Richards+2011}), we utilize a hybrid metric, the \civ\ ``distance", as defined by \cite{Rivera+20}.  This distance is relative to the best-fit curve tracing the locus of points in the \civ\ EW-blueshift plane, where quasars with large EW and small blueshift have small distance, while quasars with small EW and large blueshift have large distance.  This metric should be more robust in terms of identifying quasars with slim-disk-like geometries than EW or blueshift alone. 
Fifteen of the quasars in our SDSS-RM subsample of 133 with the most extreme \civ\ distances have blueshifts in excess of 1500\,km s$^{-1}$ and are shown in Figure~\ref{fig:fig1}; 14 of these have X-ray information.
%and are indicated with red dots in Figure~\ref{fig:fig2}.
%For example, the scale height of the slim disk may be small in our sample as compared to true WLQs.  

A logical next step would be to examine the \citet{Luo+2015} and \citet{Ni+2018} WLQs for AAL0s to determine if X-ray weakness correlates with the presence of AAL0s.  However, that experiment is impossible with the current data as both samples
excluded ``objects with narrow absorption features around \civ'' in
addition to BALs and mini-BALs.  Thus the \citet{Luo+2015} and \citet{Ni+2018} data do not enable a test of this hypothesis---despite otherwise being ideal samples.  Further analysis of AAL0 systems in WLQs is needed.

\section{Discussion and Conclusions}
\label{sec:conclusions}

While our test does not provide definitive proof that AAL0s have an edge-on orientation, our results suggest that there is merit to following up this hypothesis further.  It is important to note that our test relying on X-ray weakness as a potential indicator of edge-on orientation applies only to objects with slim-disk geometries, but the AAL0 orientation hypothesis posed is generic.  That is, we would predict that all of the AAL0 systems in Figure~\ref{fig:fig2} are observed edge-on, but do not expect the AAL0 objects with small \civ\ distances to be X-ray weak, as the standard disk geometry is not expected to hide the X-rays in such sources.    Indeed such objects (with large \civ\ EW) are consistent with relatively ``hard" SEDs and a ``failed" wind as discussed by \citet{Giustini+2019}.  The crucial point is that our test of a very small sample could potentially be leveraged into an orientation indicator for a far larger number of quasars (those with moderately high resolution and high-S/N coverage of the \civ\ emission line) than currently have orientation estimates (which currently requires sensitive radio or X-ray observations), particularly those that are radio quiet.

Additional work is needed to explore the narrow absorption-line properties of an unbiased sample of WLQs in order to determine if both results predicted by the orientation hypothesis presented herein are confirmed.  Further tests would also benefit from X-ray spectral analyses, as X-ray weak edge-on systems accreting from a slim disk should display harder X-ray spectra \citep{Ni+2018}.  A larger sample and/or time-resolved X-ray data would also be beneficial, as \citet{Ni+2020} find that X-ray weak WLQs can transform to X-ray normal; thus we should not necessarily expect that all potential slim-disk AAL0 systems be X-ray weak at any given time.

%Unfortunately, an insufficient number of AAL0 source in our sample have useful X-ray spectral information to perform a meaningful population study.

\acknowledgments

We thank Teng Liu for access to the X-ray data in advance of publication, Vivienne Wild and Joe Hennawi for discussions about the incidence of \civ\ absorption, and Trevor McCaffrey for constructing the \civ\ distance metric.  This research has made use of data obtained from the Chandra Source Catalog (https://cxc.harvard.edu/csc), provided by the Chandra X-ray Center (CXC) as part of the Chandra Data Archive.  Funding for SDSS-III has been provided by the Alfred P.\ Sloan Foundation, the Participating Institutions, the National Science Foundation, and the U.S. Department of Energy Office of Science. The SDSS-III website is http://www.sdss3.org/.

%\vspace{5mm}
%\facilities{}


%\software{          }



%\appendix

\bibliography{refs,sdssrm}
\bibliographystyle{aasjournal}



\end{document}